\documentclass[conference]{IEEEtran}
\IEEEoverridecommandlockouts
\usepackage{cite}
\usepackage{amsmath,amssymb,amsfonts}
\usepackage{algorithmic}
\usepackage{graphicx}
\usepackage{textcomp}
\usepackage{xcolor}
\usepackage{balance}

\def\BibTeX{{\rm B\kern-.05em{\sc i\kern-.025em b}\kern-.08em
    T\kern-.1667em\lower.7ex\hbox{E}\kern-.125emX}}

\begin{document}

\title{Battery-less Long-Range LTE-M \\Water Leak Detector\\

}

\author{
    \IEEEauthorblockN{Roshan Nepal\IEEEauthorrefmark{1},
    Brandon Brown\IEEEauthorrefmark{2},
    Shishangbo Yu\IEEEauthorrefmark{1},\\
    Roozbeh Abbasi\IEEEauthorrefmark{2},
    Norman Zhou\IEEEauthorrefmark{2},
    George Shaker\IEEEauthorrefmark{1}}\\
    
    \IEEEauthorblockA{\IEEEauthorrefmark{1}Dept. of Electrical and Computer Engineering, University of Waterloo, Canada\\
    \{roshan.nepal, s342yu, gshaker\}@uwaterloo.ca}
    
    \IEEEauthorblockA{\IEEEauthorrefmark{2}Dept. of Mechanical and Mechatronics Engineering, University of Waterloo, Canada\\
    \{bbrown, rabbasi, nzhou\}@uwaterloo.ca}
}

\maketitle

\begin{abstract}
This work presents a self-powered water-leak sensor that eliminates both batteries and local gateways. The design integrates a dual-compartment electrochemical harvester, a low-input boost converter with supercapacitor storage, and a comparator-gated LTE-M radio built on the Nordic Thingy:91 platform. Laboratory tests confirm that the system can be awakened from a dormant state in the presence of water, harvest sufficient energy, and issue repeated cloud beacons using the water exposure as the power source. Beyond conventional LTE-M deployments, the system’s compatibility with 3GPP-standard cellular protocols paves the way for future connectivity via non-terrestrial 5G networks, enabling coverage in infrastructure-scarce regions.
\end{abstract}


\section{Introduction}
Water leakage remains a persistent challenge across residential and industrial settings, often resulting in financial losses, operational disruptions, and infrastructure damage. In households, undetected leaks can cause structural deterioration and increased utility costs. The U.S. Environmental Protection Agency estimates that household leaks in the U.S. waste nearly one trillion gallons of water annually \cite{epa_fix_a_leak}. In industrial settings, even small leaks can disrupt operations or damage critical equipment, such as halting sterile processes in pharmaceutical plants or damaging servers in data centers \cite{kadu2015, ahopelto2020}. These risks underscore the need for autonomous, low-maintenance leak detection systems capable of timely alerts with minimal environmental footprint \cite{zaman2020, moubayed2021}.

Hydroelectric energy harvesting has emerged as a sustainable solution, leveraging water-triggered electrochemical reactions to power sensors only during leak events \cite{Rouhi2024, feng2020high}. By eliminating batteries, such systems reduce maintenance, avoid long-term dormancy issues, and prevent electronic waste. Their event-driven nature enables dormant operation with immediate responsiveness to leaks, supporting both longevity and environmental goals.

However, communication remains a bottleneck for practical deployment. Conventional battery-free leak detectors often use BLE, LoRa, or Zigbee to transmit alerts \cite{alshami2024, jan2022}. While energy-efficient, these protocols depend on nearby gateways to access cloud platforms, which can be costly and impractical in large or remote deployments \cite{jeon2018, haxhibeqiri2018}. Battery use also introduces environmental and operational burdens due to replacements and disposal \cite{hasan2023}.

To address these issues, this paper introduces a battery-free leak detection system that combines hydroelectric energy harvesting with direct LTE-M connectivity. LTE-M, a low-power cellular LPWAN standard, enables long-range cloud communication without local gateways \cite{moges2023, vaezi2022}. This architecture supports autonomous sensors capable of activating from harvested energy and transmitting alerts over existing cellular infrastructure, making it well-suited for deployment in inaccessible or maintenance-limited environments.

Experiments confirm the system's viability, with successful energy harvesting and direct cloud communication in response to leak events. The results highlight a practical path toward scalable, maintenance-free leak monitoring that combines energy autonomy with infrastructure-independent connectivity.

\section{System Architecture and Methodology}

The sensor unit is designed to harvest energy from water-triggered electrochemical reactions using a nanomaterial-based structure. It consists of a layered stack made of carbon nanofibers (CNFs) mixed with salt, placed between reactive electrodes like aluminum (Al), magnesium (Mg), etc~\cite{feng2020high}. Upon exposure to water, the system initiates a combination of galvanic and electrochemical reactions, producing electricity without requiring any external power source. This harvested energy forms the foundation for fully battery-free operation.

In earlier designs, this sensor was used in two separate systems. In the BLE-based implementation, harvested energy was used to send a low-energy broadcast packet to a smartphone or a gateway nearby~\cite{Rouhi2024}. In the LoRa-based version, a larger sensor was paired with a DC-DC boost converter and a 100~mF supercapacitor~\cite{nepal2025lora}. The system could generate enough energy to power a LoRa module and transmit to a nearby gateway over several meters. Both systems demonstrated proof-of-concept for battery-free leak detection, but their communication range was limited by protocol constraints and the need for local gateways.

\begin{figure}[!t]
\centerline{\includegraphics[width=80mm]{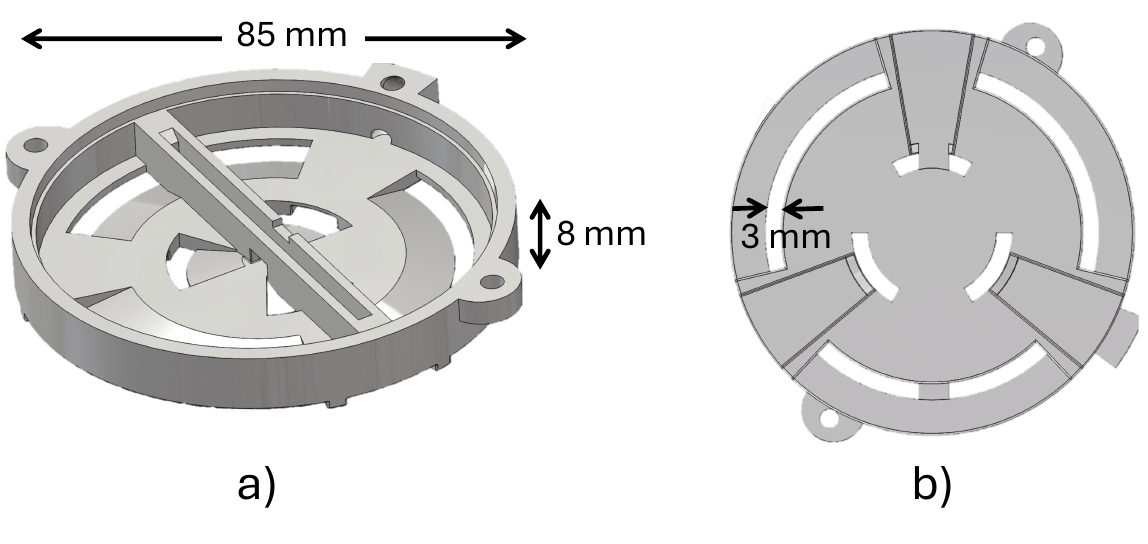}}
\caption{CAD representation of the custom sensor enclosure: (a) Isometric view highlighting the dual-compartment layout for electrode and electrolyte placement; (b) Bottom view showing integrated inlet pathways designed to facilitate water ingress and initiate electrochemical energy generation.}
\label{fig:sensor_unit}
\end{figure}

The LTE-M communication system is built around the Nordic Thingy:91 evaluation board, which integrates the nRF9160 SiP, a low-power LTE-M/NB-IoT modem. LTE-M modules like this require a minimum operating voltage of 3.2V and can draw peak currents of up to 250mA during transmission~\cite{nordic2022thingy91}. To reduce the burden on the boost converter and avoid excessive current losses during voltage stepping, a compartmentalized sensor design was adopted. As shown in Fig.~\ref{fig:sensor_unit}, two electrochemical cells with identical material compositions are connected in series inside a custom enclosure. This configuration increases the open-circuit voltage (OCV) without significantly affecting the short-circuit current (SCC). This makes the sensor output better suited for powering higher-demand modules like the nRF9160.

\begin{figure}[!t]
\centerline{\includegraphics[width=90mm]{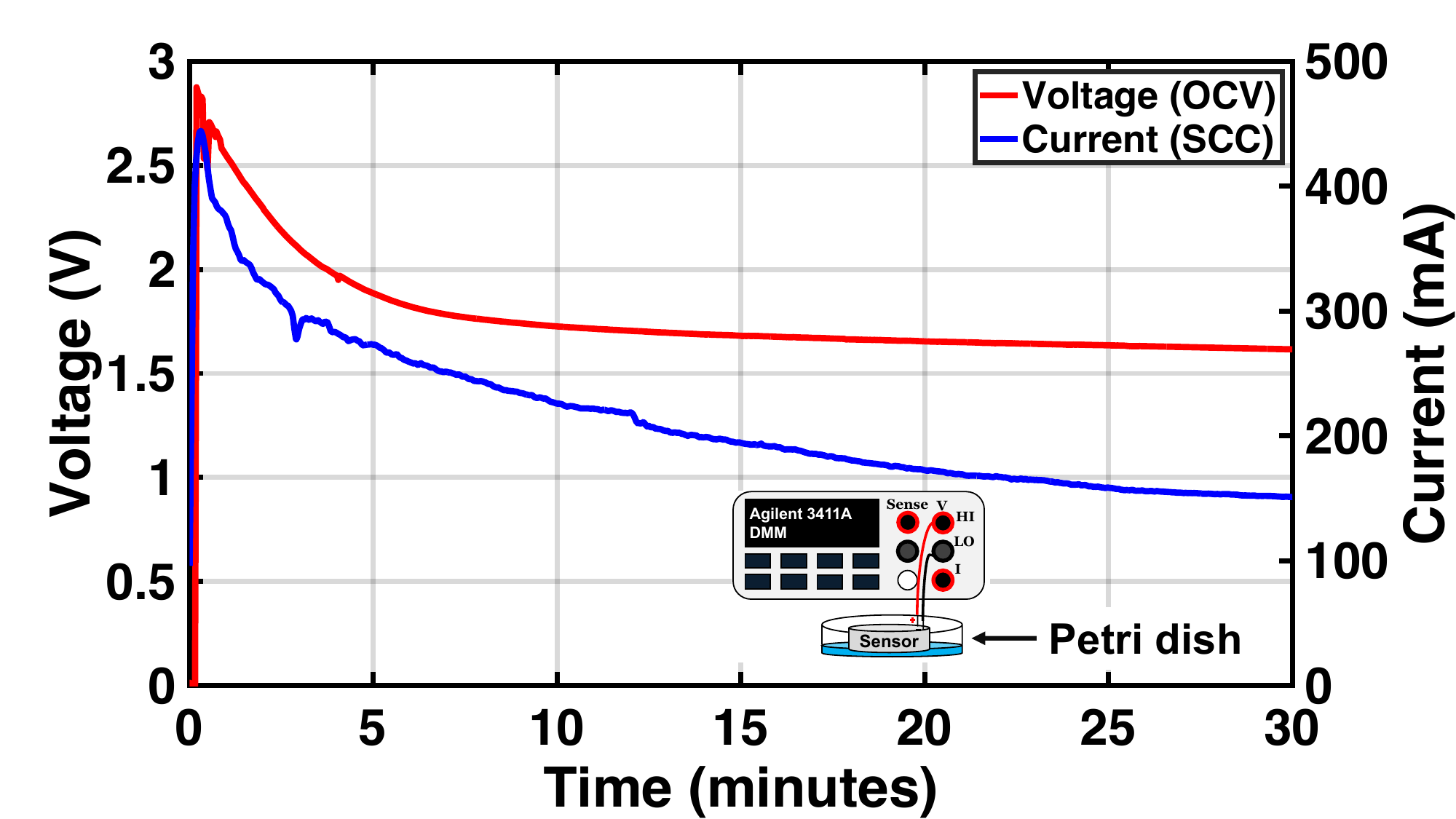}}
\caption{OCV (left y-axis) and SCC (right y-axis) output of the sensor over time after water exposure.}
\label{fig:OCV_SCC}
\end{figure}

To evaluate the performance of the dual-compartment sensor, we measured its OCV and SCC over time following exposure to water in a 1~mm-deep Petri dish. This shallow water depth was chosen to simulate realistic early-stage leaks. Measurements were taken using an Agilent 3411A digital multimeter~\cite{keysight34410A34411A}, with results shown in Fig.~\ref{fig:OCV_SCC}. The sensor produces a peak OCV of around 2.7~V shortly after activation, stabilizing near 1.6~V within 30 minutes. The SCC starts above 450~mA and levels out at approximately 150~mA.

The harvested energy is stepped up by a DC-DC boost converter based on the ME2108 IC and stored in a 1.5~F supercapacitor, selected based on the system’s energy availability and load requirements. This capacitor serves as an energy buffer and is sized to handle the entire LTE-M wake-up and transmission sequence. While a smaller capacitor would reduce charging time, it would not reliably sustain the modem through its peak current demands. A 1.5~F value was chosen to balance energy availability and recharge duration.

\begin{figure}[!t]
\centerline{\includegraphics[width=80mm]{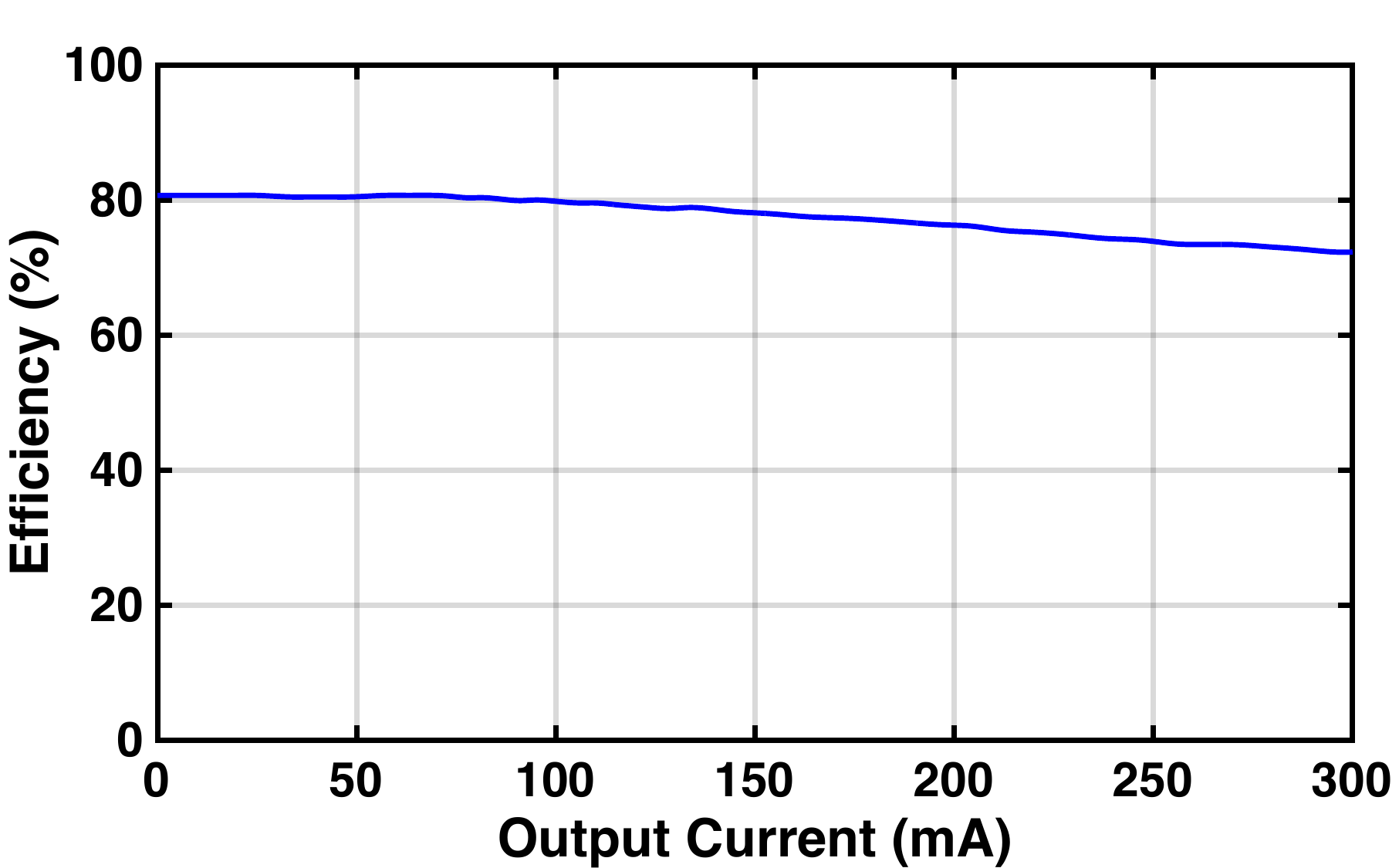}}
    \caption{Measured efficiency of the 5~V boost converter at 1.2~V input with varying output currents.}

\label{fig:boost_efficiency}
\end{figure}

The ME2108 offers reliable voltage conversion performance even at low input voltages, making it particularly suitable for energy-harvesting applications. As shown in Fig.~\ref{fig:boost_efficiency}, the converter maintains efficiency above 80 \% under light load conditions and delivers approximately 73 \% efficiency at the peak 250~mA load required during LTE-M startup. While not the most efficient converter on the market, its ability to start from input voltages as low as 0.9~V and operate down to 0.5~V gives it a distinct advantage in systems where the harvested voltage frequently drops below the thresholds of conventional converters. This trade-off ensures consistent performance and uninterrupted energy delivery, even under fluctuating sensor output conditions.

\begin{figure*}[!t]
\centerline{\includegraphics[width=140mm]{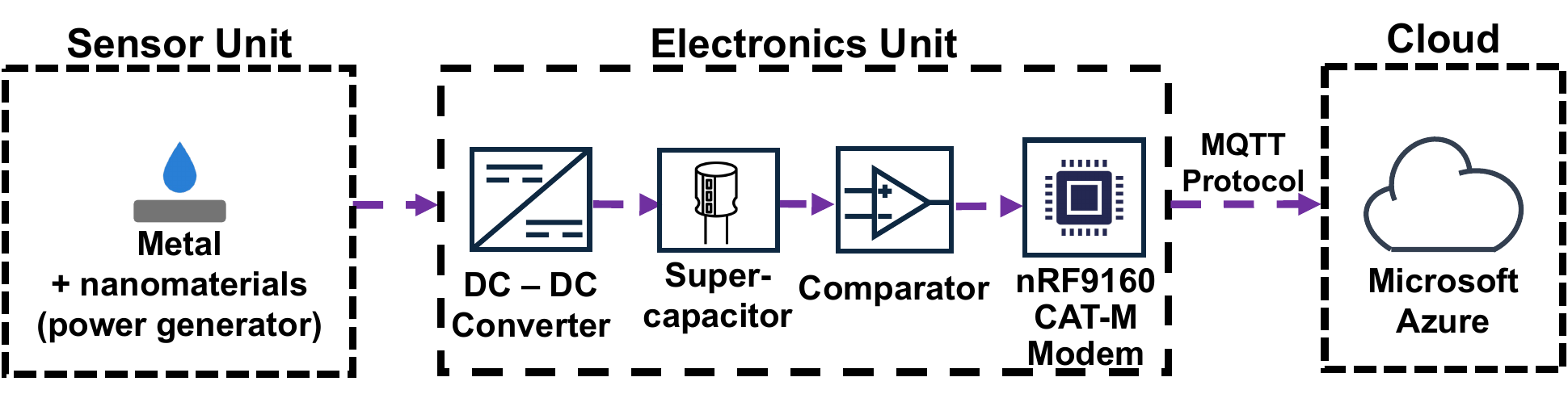}}
\caption{System architecture of the self-powered LTE-M system, where harvested energy powers the nRF9160 modem to transmit leak data to the cloud via MQTT.}
\label{fig:system_architecture}
\end{figure*}

One key design challenge is managing the power flow to the LTE-M module. The Nordic Thingy:91 begins drawing current immediately upon connection, which can lead to premature discharge of the supercapacitor if it has not yet reached a sufficient voltage. To address this, we use a TLV431-based comparator circuit to isolate the load. The comparator monitors the supercapacitor voltage and switches the load only when it exceeds a predefined threshold (e.g., 4.87~V), ensuring that enough energy is available for reliable transmission. The voltage thresholds are set using a resistor divider network, which also introduces hysteresis, disconnecting the LTE-M module once the voltage drops below a lower limit (e.g., 3.67~V). This ensures stable cycling between energy accumulation and communication, avoiding premature brownout conditions. 

Once the supercapacitor voltage surpasses the comparator’s turn-on threshold, the switch closes to power the Nordic Thingy:91 module. The module then initiates the LTE-M network attach sequence, searches for a valid cellular signal, and upon successful connection publishes a MQTT message to Azure IoT Hub. Each payload is formatted as a JSON object containing the \texttt{device\_id} and  a \texttt{timestamp}, and is approximately 50 bytes in size.

Fig.~\ref{fig:system_architecture} shows the overall block diagram of the system, consisting of the dual-compartment sensor, boost converter, supercapacitor, comparator switch, and LTE-M modem. The system operates fully on harvested energy and performs direct cloud communication without requiring batteries or local gateways.

\section{Experimental Validation}
To validate the end-to-end functionality of the proposed system, we conducted closed-loop experiments incorporating the sensor, power management unit, supercapacitor, and LTE-M communication module. The primary metric of interest was the voltage across the supercapacitor, as it reflects the system’s ability to harvest, store, and deliver energy autonomously for data transmission.

The test was performed under room temperature conditions using a Petri dish containing a 0.5~mm water layer, intended to replicate early-stage leak scenarios. Upon water exposure, the electrochemical reaction began, initiating capacitor charging. The voltage was monitored continuously using a digital multimeter.

\begin{figure}[!t]
\centerline{\includegraphics[width=80mm]{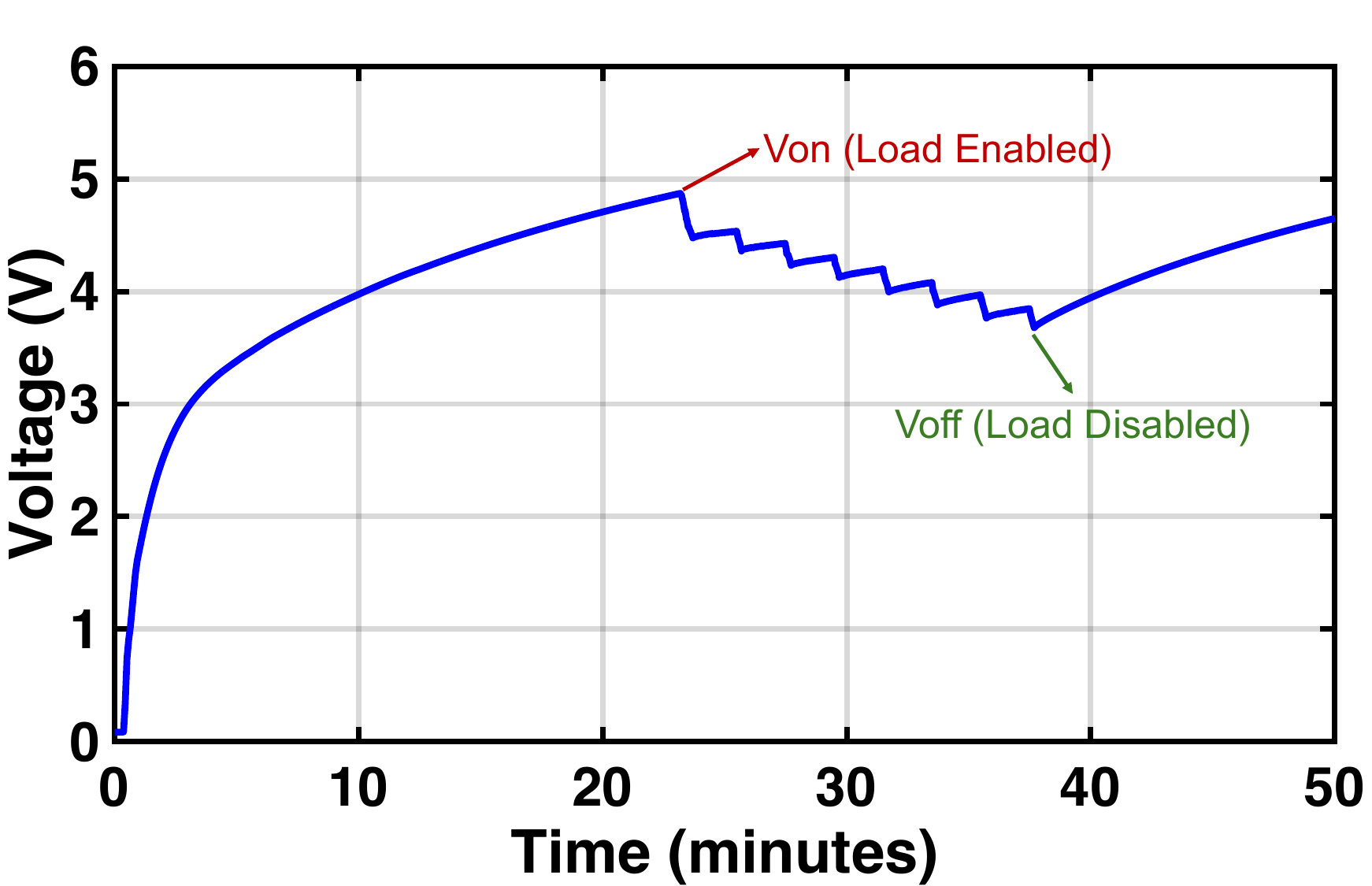}}
    \caption{Supercapacitor voltage profile showing comparator-based load control. The load activates at \(V_{\text{on}} = 4.87~V\) and deactivates at \(V_{\text{off}} = 3.67~V\), with periodic voltage dips marking successive beacon transmissions at 2-minute intervals.}
\label{fig:comparator_behavior}
\end{figure}

As shown in Fig.~\ref{fig:comparator_behavior}, the supercapacitor reached the activation threshold of approximately 4.87~V after 23 minutes. At this point, the comparator circuit enabled the load, powering the Thingy:91 module. This was followed by a distinct voltage dip, corresponding to the module’s network acquisition and the first LTE-M beacon transmission. Based on ten trials conducted across diverse indoor and outdoor environments, the network acquisition time—measured from the moment the load is enabled to successful LTE-M attachment—ranged between 10 and 20 seconds, with an average of approximately 14 seconds. This variation primarily depended on signal availability and proximity to the nearest cellular base station. Despite the high current demand estimated at 250~mA, no brownout was observed, confirming that the energy buffer and isolation mechanism operated as intended.

The Thingy:91 module was programmed to enter a 2-minute idle phase before attempting the next beacon. This phase allows the 1.5 F supercapacitor to recharge energy, so each transmission starts from a higher voltage and the system can sustain more beacons on a single charge. In practice, the 2-minute window also acts as a secondary, time-based gate that complements the TLV431 voltage thresholds, preventing premature brownouts and maximizing the number of successful pings per cycle. As shown in Fig.~\ref{fig:comparator_behavior}, the capacitor voltage rises during each idle interval, with periodic dips every 2 minutes marking successive beacon events. The charge–transmit–recharge sequence continues until the voltage falls to approximately 3.67 V, at which point the comparator disconnects the LTE-M load and the capacitor resumes uninterrupted charging. While the 2-minute interval offered reliable performance in our experiments, it is fully configurable in firmware—longer idle durations can be used to enable more beacon transmissions per charge cycle when needed.

A single charge–discharge cycle of the 1.5~F capacitor supported eight consecutive beacons on average with 2-minute idle window before isolation occurred. Moreover, since the transmission uses the existing LTE-M cellular network, geographic range is dictated by cellular coverage rather than any additional hardware constraints.

This result confirms that the system can autonomously manage high-power communication cycles using only harvested energy. The consistent voltage profile and controlled load activation demonstrate effective integration of sensing, power regulation, and communication in a battery-free configuration. Once the leak subsides and the sensor dries, the electrochemical reaction becomes highly inefficient. While reapplying water triggers a limited response, the regenerated output is minimal and inconsistent due to non-uniform redistribution and changes in the electrode surface. As a result, the sensor effectively behaves as a single-use device under the current design. Reactivation would require refurbishment or replacement of the sensor to restore consistent performance. However, given that water leak events are typically infrequent and often trigger inspection or maintenance procedures, the water sensor component can be conveniently replaced as part of the standard maintenance workflow (without need for replacing any of the underlying electronics).

\section{Conclusion}
This work presents a novel battery-free water leak detection system that combines hydroelectric energy harvesting with direct LTE-M communication. Through a compartmentalized sensor architecture and low-voltage power management design, the system achieves reliable activation and cloud-based data transmission without batteries, external power, or local gateways. Experimental validation under controlled leak conditions confirmed that harvested energy alone is sufficient to support multiple LTE-M beacon transmissions per charge cycle. Moreover, since the system uses a 3GPP-compliant LTE-M modem, it is compatible with both terrestrial and emerging non-terrestrial 5G networks, enabling deployment in remote or infrastructure-limited environments. These findings demonstrate the feasibility of scalable, battery-free, low-maintenance monitoring systems that can autonomously operate in hard-to-access locations with minimal infrastructure overhead.

\section*{Acknowledgment}
The authors would like to acknowledge the support from MITACS, Rogers Communications, and AquaSensing Inc. 

\balance 

\bibliographystyle{IEEEtran}
\bibliography{references}

\end{document}